\documentclass[superscriptaddress,twocolumn,aps,showpacs,pre]{revtex4-1}

\pdfoutput = 1
\usepackage[english]{babel}

\keywords{Bose-Einstein-Condensation,Critical Phenomena}

\usepackage{amsmath}
\usepackage{epstopdf}
\usepackage{setspace}

\usepackage{xcolor}     
\usepackage{soul}       

\begin{document}

\title{Non-classical critical exponents at ideal Bose-Einstein condensation}

\author{I. Reyes-Ayala, F. J. Poveda-Cuevas, and V. Romero-Roch\'{i}n}
\affiliation{Instituto de F\'{i}sica, Universidad Nacional Aut\'{o}noma de M\'{e}xico, Apartado Postal 20-364, 
01000 Ciudad de M\'{e}xico, Mexico}

\email{romero@fisica.unam.mx}

\date{\today}

\begin{abstract}
We show that ideal Bose-Einstein condensation (BEC) in $d = 3$ dimensions is a non-classical critical second order phase transition with exponents $\alpha = -1$, $\beta = 1$, $\gamma = 1$, $\delta = 2$, $\eta = 1$ and $\nu = 1$,  obeying all the scaling equalities. These results are found with no approximations or assumptions. The previous exponents are a critical universality class on its own, different from the so-far accepted notion that BEC belongs to the Spherical Model universality class.
\end{abstract}

\maketitle

\section{Introduction}

Bose-Einstein Condensation (BEC) \cite{Einstein1925} of an ideal Bose gas is a trascendental physical problem of our time. Despite being an {\it ideal} gas, namely, without interatomic interactions, it has opened a path to our understanding of macroscopic matter at the quantum regime, typically at very low temperatures. While it is not our purpose to review it here, we cannot avoid mentioning the extraordinary experimental advances of the last decades\cite{Anderson:1995aa,Davis:1995aa,Dalfovo:1999aa,Bloch:2000aa,Regal:2004aa,Dalibard:2011aa} motivated by BEC and leading to a vigorous field of theoretical and experimental research now known as quantum matter. As one would imagine, ideal BEC being now a well understood textbook topic, there should not be anything ``novel'' about it. However, as we show here, ideal BEC, accepted as a bona-fide second order phase transition, has been considered in a wrong universality class. That is, instead of belonging to the Spherical Model (SM) universality class,\cite{Berlin:1952aa,Gunton-prl66,Hall:1975aa,Baldo1976,Caldi1981,Rasolt1984,Xian-Zhi1999,Search2001,DianaV.Shopova2002,Jakubczyk:2013aa} BEC has its own universality class, a non mean-field one, with a set of non-classical exponents. The assessment of ideal BEC belonging to SM universality class resides in the work by Gunton and Buckingham (GB) \cite{Gunton-prl66}, in which it is first {\it assumed} that the order parameter is the condensate particle wavefunction, and then, the transition is considered within a mean-field approximation.
Here without any assumption, but simply by following the exact results of the usual ideal Bose gas, we show that BEC transition is driven by the particle density and, with no approximations, exhibits non-classical exponents in the behavior of the usual thermodynamic variables, pressure $p$, density $\rho$, isothermal compressibility $\kappa_T$ and in the corresponding density-density correlation function. As a consequence, these findings {\it suggest} that the proper order parameter is the condensate particle {\it density}.  It is of interest to highlight here that the non-classical critical BEC behavior, as the Ising model in $d =2$, are among those rare cases where the $d = 3$ exponents can be found exactly. \\

We will proceed as straightforwardly as possible. First, we briefly review the phenomenon of Bose-Einstein condensation in order to make our study as clear as possible; then, we stress out the presence of critical density fluctuations at BEC, the landmark of critical phenomena, and calculate the exponents $\gamma$, $\eta$ and $\nu$ of the isothermal compressibility, of the correlation length and of the correlation function at criticality, respectively. Next, we calculate the exponents $\alpha$ and $\delta$, the first one indicating the behavior of the heat capacity at constant volume, and the second one regarding the approach to criticality of the pressure as a function of the density. All the previous exponents will be calculated as the transition is approached from above criticality. Then, we present the essential details of the work by Gunton and Buckingham \cite{Gunton-prl66}  that lead to the so-far accepted conclusion that ideal BEC critical behavior belongs to the Spherical Model class. We shall show that such an approach is not only of approximated validity but that it is of a mean-field character, showing no fluctuations of the assumed order parameter.
Following these results we will then argue that the behavior above criticality suggests that the order parameter of the transition is the condensate fraction and not the condensate wavefunction. This will lead to the exponent $\beta$ and to the verification that the exponents obey the scaling critical equalities. We shall conclude with a brief discussion of the behavior of the gas within the BEC phase and argue for the stability of the thermal gas below BEC.

\section{Ideal Bose-Einstein Condensation} 

We consider an ideal $d = 3$ gas of Bose particles of spin $s = 0$ and mass $m$, contained in a vessel of volume $V$. The Grand Canonical partition function is,
\begin{equation}
\Xi = {\rm Tr} \exp[{- (\hat H - \mu \hat N)/kT}] \label{Xi0}
\end{equation}
with $T$ and $\mu$ the temperature and chemical potential, $k$ Boltzmann constant, $
\hat H = \sum_{\vec p} \epsilon_p \hat a_{\vec p}^\dagger \hat a_{\vec p}$,
the ideal gas Hamiltonian, where $\epsilon_p = p^2/2m$,  $\hat a_{\vec p}$ the annihilation operators of particles with momentum $\vec p$, and $\hat N = \sum_{\vec p} \hat a_{\vec p}^\dagger \hat a_{\vec p}$ the number operator. In the thermodynamic limit the Grand Potential $\Omega = -kT \ln \Xi$ can be obtained,\cite{Landau1980statistical}
\begin{equation}
\Omega(V,T,\mu) = - kT \frac{V}{\lambda_T^3} g_{5/2}(\mu/kT), \label{Omega}
\end{equation}
where $\lambda_T = h/\sqrt{2 \pi m kT}$ is de Broglie thermal wavelength and $g_{5/2}(\mu/kT)$ the $s =5/2$ Bose function,
\begin{equation}
g_{s}(\alpha) = \frac{1}{\Gamma(s)} \int_0^\infty \frac{x^{s-1} dx}{e^{x - \alpha} -1} .\label{Boseg}
\end{equation}
These functions are defined for $\alpha \le 0$ only, being non-analytic at $\alpha = 0$. In particular, for fractional $s$, the Bose functions obey the following asymptotic expression for $|\alpha| < 2\pi$, \cite{Wood1992},
\begin{equation}
g_s(\alpha) = \Gamma(1-s) |\alpha|^{s-1} + \sum_{k =0}^{\infty} \frac{\zeta(s-k)}{k!} \alpha^k .\label{asin}
\end{equation}
For $s = 5/2$, one finds that the term $\sim |\mu|^{3/2}$ in $\Omega$, gives rise to an essential singularity at $\mu = 0$, which in turn, rules the critical behavior at BEC, as discussed below. This singularity will be contrasted below with the mean-field approximation that incorrectly yields that BEC is in the SM universality class.\\

From the grand potential one can obtain all the thermodynamics of the gas. \cite{Landau1980statistical} Let us first look at the number of particles of the gas, 
\begin{equation}
N = \frac{V}{\lambda_T^3} g_{3/2}(\mu/kT). \label{N}
\end{equation}
When the chemical potential vanishes, $\mu = 0$, the gas reaches a state with a temperature $T_c$, which is a function of the homogenous density $\rho = N/V$, namely,
\begin{equation}
\rho = \frac{1}{\lambda_{T_c}^3} \zeta(3/2), \label{Nc}
\end{equation}
with $g_{3/2}(0) = \zeta(3/2)$ the Riemann-zeta function evaluated at 3/2. Below such a temperature, for a given density $\rho$, BEC occurs, namely, the gas enters a different gas phase where a macroscopic number of atoms occupy the zero momentum state $\vec p = 0$. In this phase the chemical potential remains zero, $\mu = 0$, all the way down to $T = 0$. To be more precise, we recall that in equilibrium the number of atoms in the one particle state $\vec p$ is given by the Bose-Einstein distribution,
\begin{equation}
\overline{n}_{\vec p} = \frac{1}{e^{(\epsilon_p - \mu)/kT}-1} \label{BE-dis} .
\end{equation}
Hence, in the thermodynamic limit, for $T > T_c$, $\mu < 0$, the density of particles in the ground state $\vec p = 0$, is zero, in the sense that $\overline n_0/V = 0$, as $V \to \infty$. Now, for $T \le T_c$, the chemical potential remains zero, and while the occupancy number $\overline n_{\vec p}$ is still given by Eq.(\ref{BE-dis}) for $\vec p \ne 0$, the density of particles in the  ``condensate'' is now given by,
\begin{equation}
\frac{\overline n_0}{V} = \rho\left( 1 - \left(\frac{T}{T_c}\right)^{3/2}\right).\label{condensate}
\end{equation}
As BEC is approached from above, $\mu \to 0^-$, $T \to T_c^+$, and from expressions Eqs. (\ref{asin}), (\ref{N}) and (\ref{Nc}) one can find how $\mu$ depends on $|T - T_c|$, for a given density $\rho $, 
\begin{equation}
|\mu|^{1/2} \approx \frac{3}{4\sqrt{\pi}} \zeta(3/2) (kT_c)^{1/2} \left( \frac{T-T_c}{T_c} \right).\label{mu0}
\end{equation}

\section{Critical density fluctuations, isothermal compressibility and correlation length} 

The signature of critical phenomena, in a fluid, resides on the fact that the density-density correlations show critical fluctuations. \cite{Landau1980statistical,Fisher:1967aa,ma2018modern} These fluctuations are related to the divergence of the isothermal compressibility $\kappa_T$ at $T_c$. This thermodynamic quantity, for $\mu < 0$, is given by
\begin{eqnarray}
\kappa_T & = & \frac{1}{\rho^2} \left(\frac{\partial \rho}{\partial \mu} \right)_{T} \nonumber \\
& = & \frac{1}{kT} \left(\frac{1}{\rho\lambda_T}\right)^2 \> g_{1/2}(\mu/kT) .\label{kappa}
\end{eqnarray}
Using the asymptotic form of $g_{1/2}(\mu/kT)$ as given by Eq.(\ref{asin}), and that of $\mu$, Eq.(\ref{mu0}),  we find that for fixed density $\rho$, as $T \to T_c^+$,
\begin{equation}
\kappa_T \approx \frac{4 \pi }{3\zeta(3/2)} \frac{1}{kT_c} \left(\frac{1}{\rho \lambda_T}\right)^2 \left( \frac{T-T_c}{T_c} \right)^{-1} . \label{gamma}
\end{equation}
This expression indicates that the isothermal compressibility diverges at $T_c$, as $\kappa_T \sim (T-T_c)^{-1}$. This defines the critical exponent $\gamma$,\cite{Fisher:1967aa,ma2018modern} being $\gamma = 1$.\\

The density-density correlation function is given by, \cite{Landau1980statistical,Ziff:1977aa}
\begin{eqnarray}
G(\vec r - \vec r^\prime) & = & \langle \hat \rho(\vec r) \hat \rho(\vec r^\prime) \rangle - \langle \hat \rho(\vec r) \rangle \langle \hat \rho(\vec r^\prime) \rangle \noindent \\
& = & \rho \delta(\vec r - \vec r^\prime) + \left|\frac{1}{h^3}  \int \frac{e^{i\vec p\cdot \vec r/\hbar} d^3p}{e^{\beta\left(\frac{p^2}{2m}-\mu\right)}-1}\right|^2 \nonumber
\end{eqnarray}
where $\hat \rho(\vec r) =  \hat \Psi^\dagger(\vec r) \hat \Psi(\vec r)$ is the particle density operator with $\hat \Psi(\vec r)$ the annihilation particle operator at spatial point $\vec r$. The relationship between the density fluctuations and the isothermal compressibility is obeyed, 
$\int G(\vec r) d^3 r = \rho^2 kT \kappa_T$. Near BEC,  as $\mu \to 0$, it can be shown \cite{Ziff:1977aa} that the term 
responsible in $G(\vec r)$ for the critical fluctuations is given by
\begin{equation}
G_c(\vec r)  \approx \frac{1}{2\lambda_T^4}\frac{e^{-r/\xi}}{r^2} \label{gc}
\end{equation}
where the correlation length is $\xi = \hbar/2 \sqrt{2 m|\mu|}$. Therefore, as $\mu \to 0$, for $\rho=N/V$ constant and using Eq. (\ref{mu0}), the correlation length diverges as
\begin{equation}
\xi \sim \> \left( \frac{T-T_c}{T_c} \right)^{-1} ,
\end{equation}
thus defining the exponent $\nu$\cite{Fisher:1967aa,ma2018modern}, as $\nu = 1$. 
It is worthwhile to point out that, since there are no atomic interactions, all particle correlation functions decay with a correlation length proportional to $\xi$, see Ref. \cite{Ziff:1977aa}, indicating the essential role of the chemical potential in driving BEC.\\

At criticality, $\mu = 0$, the correlation function becomes long range, scaling as $G(r) \sim 1/r^2$, which when compared with the definition of the critical exponent $\eta$, \cite{Fisher:1967aa,ma2018modern} $G(r) \sim 1/r^{d-2+\eta}$, yields $\eta = 1$. It is very important to highlight here that this identification signals that BEC is a non-classical, or non mean-field, critical transition: the function $G(r)$ cannot be expressed in the Orstein-Zernike form \cite{Fisher:1967aa,ma2018modern}. We return to this point below.

\section{Heat capacity and critical isotherms} 

The heat capacity at constant volume is given by,
\begin{eqnarray}
C_V& = & T \left(\frac{\partial S}{\partial T}\right)_{N,V} \noindent \\
& = & \frac{3}{2}Nk \left[\frac{5}{2} g_{5/2}(\mu/kT) - \frac{3}{2} \frac{g_{3/2}^2(\mu/kT)}{g_{1/2}(\mu/kT)}\right] .
\end{eqnarray}
While the heat capacity is not analytic at $\mu = 0$, it does not diverge. Instead, it shows a ``peak'' at BEC where its first derivative is not continuous. This allows to define a critical exponent that turns out to be negative, similarly as it does occur in the $d=3$ XY model \cite{Lipa:2003aa,Burovski:2006aa}. The behavior near $\mu = 0$, in terms of the critical temperature can be shown to be,
\begin{equation}
C_V - C_0 \approx \frac{3}{2} Nk \left( \frac{15}{8}\zeta(5/2) - \frac{27}{8\pi}\zeta^3(3/2)\right) \left( \frac{T-T_c}{T_c} \right),
\end{equation}
where $C_0 = (15 Nk/4)\zeta(5/2)$ is the heat capacity at criticality. From the above expression, we can identify the exponent $C_V \sim |T-T_c|^{-\alpha}$ \cite{Fisher:1967aa,ma2018modern} as given by $\alpha = -1$.\\

Since BEC can occur for any temperature $T$ given a value of the density $\rho = N/V$, or molar volume $v = V/N$, all isotherms are ``critical''. Hence, the interest is in expressing the pressure difference $p - p_c$ as a function of $v - v_c$, or $\mu$ as a function of $\rho-\rho_c$, along any isotherm, as
\begin{equation}
p - p_c = p_c\left(\frac{g_{5/2}(\mu/kT)}{\zeta(5/2)} - 1\right) 
\end{equation}
\begin{equation}
\rho - \rho_c= \rho_c \left(\frac{g_{3/2}(\mu/kT)}{\zeta(3/2)} - 1\right)
\end{equation}
with $p_c$ and $\rho_c = 1/v_c$ the critical values at BEC for the given isotherm, $p_c = (kT/\lambda_T^3)\zeta(5/2)$ and $\rho_c = (1/\lambda_T^3)\zeta(3/2)$. Using the expression in Eq.(\ref{asin}) for the Bose functions, one finds that near $\mu = 0$,
\begin{equation}
\frac{p - p_c}{p_c} \approx\frac{\zeta^3(3/2)}{4\pi \zeta(5/2)} \left(\frac{v-v_c}{v_c}\right)^2
\end{equation}
or alternatively, using Eq. (\ref{asin}) again,
\begin{equation}
\mu \approx - \frac{kT}{\Gamma^2(-1/2)}\left(\frac{\rho - \rho_c}{\rho_c}\right)^2 
\end{equation}
which yield, in both cases as it should, the exponent $\delta$, \cite{Fisher:1967aa,ma2018modern} given by $\delta = 2$.

\section{The spherical model approach to BEC} 

So far, we have simply calculated the expected dependence among the thermodynamic variables in the vicinity of criticality, thus identifying the corresponding critical exponents, but without expressing the nature of the transition. In particular, one should find the order parameter of the transition. In many discussions it is assumed that the ideal BEC order parameter is the condensate {\it wavefunction}, \cite{Gunton-prl66} in close analogy with the superfluid transition in $^4$He, \cite{Fisher:1967aa,Lipa:2003aa} and now in ultracold alkali gases, \cite{Anderson:1995aa,Davis:1995aa,Dalfovo:1999aa} despite the fundamental difference that the superfluid state requires necessarily the presence of interatomic interactions. Those transitions belong to the universality class of the XY-model, which deals with a two-component order parameter. In accordance with such a premise, 
in studying the ideal Bose gas, Gunton and Buckingham (GB) \cite{Gunton-prl66} postulate that the BEC order parameter is the condensate wavefunction $\Psi_0$ and, with a set of assumptions, they find that ideal BEC belongs to the Spherical Model universality class. As we now show by reviewing GB argument, their result is of a mean-field character, leading to different conclusions from those of the previous sections.
In the section following this one, we argue that it appears more natural that the order parameter is the condensate density $\overline n_0/V$ rather than the wavefunction. Furthermore, we shall discuss the exponent equalities and the emergence of ideal BEC belonging to its unique universality class.\\

In the GB study, the condensate wavefunction is identified as the equilibrium thermal average of the condensate particle operator,  $\hat \psi_0 = V^{-1/2} \hat a_0$,   
\begin{equation}
\Psi_0 =  \langle \hat \psi_0\rangle
\end{equation}
with $\hat a_0$ the creation operator of particles with momentum $\vec p = 0$. Since in the usual approach, see Eq.(\ref{Xi0}), $\Psi_0 = 0$ identically, a {\it Bose field} $\nu_0$ must be introduced to avoid its vanishing and break the symmetry. While this method appears similar to the quasi-average procedure of Bogoliubov \cite{Bogoliubov1970,Wreszinski-thmathphys194}, it is subtly different since the $\nu_0$ field not only can take {\it any} value, but it is the conjugate intensive thermodynamic variable to the assumed extensive condensate wavefunction $V \Psi_0$. In this way, all the thermodynamics of the BEC transition is solely discussed in terms of the pair $(\Psi_0,\nu_0)$, with $\nu_0=0$ being the Bose field critical value.
GB introduce $\nu_0$ in the Grand Canonical partition function as,
\begin{equation}
\Xi_{GB} = {\rm Tr} \exp[{- [ \hat H - \mu \hat N - V(\nu_0 \hat \psi_0 + \nu_0^* \hat \psi_0^\dagger)}]/kT] \label{Xi1}
\end{equation}
with $\hat H$ and $\hat N$ as given before. In the thermodynamic limit the corresponding Grand Potential can be obtained,
\begin{equation}
\Omega_{GB}(T,V,\mu,\nu_0) =  V \frac{\nu^*_0\nu_0}{\mu} + \Omega(T,V,\mu) \label{omeGB}
\end{equation}
where $\Omega(T,V,\mu)$ is the usual Bose grand potential, given by Eq.(\ref{Omega}). The first term, proportional to (modulus square of)  the Bose field $\nu_0$, and to the inverse of $\mu$ allows to find that the phase transition, $\Psi_0 = 0$ for $T > T_c$ and $\Psi_0 \ne 0$ for $T < T_c$, belongs to the universality class of the Spherical Model. \cite{Berlin:1952aa} However, as we show below, $\Omega_{GB}(T,V,\mu,\nu_0)$ given by Eq.(\ref{omeGB}) is valid for $\mu \to 0$ only and, further, it is of a mean-field nature.
The validity at $\mu \to 0$ only yields that the term $1/\mu$ is the leading singularity in $\Omega_{GB}$, while in the usual exact approach, the singularity in $\Omega$ arises from the term $|\mu|^{3/2}$, as already discussed, see Eq.(\ref{asin}). This shift in the singularity changes the universality class from its true exact values to the approximated mean-field SM.\\

Gunton and Buckingham proceed by obtaining all thermodynamics from $d\Omega_{GB} = - SdT -p dV - N d\mu - V \Psi_0 d\nu_0 - V \Psi_0^* d\nu_0*$. The condensate wavefunction is found to be,
\begin{equation}
\Psi_0 = - \frac{\nu_0^*}{\mu} .\label{psio}
\end{equation}
This equation indicates that if $\nu_0 \ne 0$, then $\Psi_0 \ne 0$. The interesting result of this approach is that, if $T \le T_c$, then  both $\nu_0 = 0$ and $\mu = 0$, and then $\Psi_0 \ne 0$, namely, giving rise to BEC. Now, using $\rho = N/V$, the density of particles is
\begin{equation}
\rho  =  \frac{\nu_0\nu_0^*}{\mu^2} + \frac{1}{\lambda_T^3}g_{3/2}(\mu/kT) ,\label{rho}
\end{equation}
and combination of the previous two equations yields,
\begin{equation}
\rho  =  \Psi_0^*\Psi_0 + \frac{1}{\lambda_T^3}g_{3/2}(\mu/kT) .\label{rhoGB}
\end{equation}
This expression indicates that $\Psi_0^*\Psi_0 = |\Psi_0|^2$ must be interpreted as the condensate particle density, since the second term in (\ref{rhoGB}) does not take into account particles with $\vec p = 0$. However, by calculating the thermal averages directly from the grand canonical density matrix, one can show that the above results are of a mean-field kind
as seen 
below. In the following expressions, the average of an operator $\hat A$ is taken as,
\begin{equation}
\langle \hat A \rangle = \frac{1}{\Xi} {\rm Tr} \left(\hat A \exp[{-\beta [ \hat H - \mu \hat N - V(\nu_0 \hat \psi_0 + \nu_0^* \hat \psi_0^\dagger)}] \right) \label{expec-value}
\end{equation}

First, we calculate the condensate wavefunction,
\begin{eqnarray}
\Psi_0 & = & \langle \hat \psi_0 \rangle \nonumber \\
& = & - \frac{\nu_0^*}{\mu} ,
\end{eqnarray}
which agrees with Eq.(\ref{psio}). The issue at hand arises, however, when we calculate the density number of particles directly,
\begin{eqnarray}
\rho & = & \frac{1}{V} \langle \hat N \rangle \nonumber \\
& = & \frac{1}{V} \langle \hat a_0^\dagger a_0 \rangle + \frac{1}{V}\sum_{\vec p \ne 0} \langle \hat a_{\vec p}^\dagger \hat a_{\vec p} \rangle \nonumber \\
& = &  \frac{1}{V} \langle \hat a_0^\dagger a_0 \rangle + \frac{1}{\lambda_T^3}g_{3/2}(\mu/kT) .\label{den-ave}
\end{eqnarray}
This expression must be compared with Eq.(\ref{rhoGB}): in order to agree, it must be true that
$\langle \hat a_0^\dagger a_0 \rangle/V = \Psi_0^* \Psi_0$. But since $\langle \hat a_0^\dagger a_0 \rangle/V = \langle \hat \psi_0^\dagger \hat \psi_0 \rangle$, what it is being found is that  $\langle \hat \psi_0^\dagger \hat \psi_0 \rangle  = \Psi_0^* \Psi_0 = \langle \hat \psi_0^\dagger \rangle \langle \hat \psi_0 \rangle$. Namely, that the order parameter shows no fluctuations, effectively enforcing a mean-field description of the transition.
 This can also be found by calculating the expectation value $\langle \hat \psi_0^\dagger \hat \psi_0 \rangle$ directly from its definition, using Eq. (\ref{expec-value}); see Ref. \cite{Wreszinski-thmathphys194} for a rigorous discussion of this last point.\\

 Now we show that GB approach is valid for $\mu \to 0$ only. For this we calculate the Bose susceptiblity, $\chi = kT (\partial \Psi_0 /\partial \nu_0^*)_T$, which again, can be calculated both from the thermodynamic identities and from its statistical mechanics expression.  For $T > T_c$, using the thermodynamic result (\ref{psio}), one finds,\begin{equation}
\chi = - \frac{kT}{\mu}  ,\label{chi1} 
\end{equation}
 and using the density matrix directly,
\begin{eqnarray}
\chi & = & \langle \hat a_0^\dagger \hat a_0 \rangle - \langle \hat a_0^\dagger\rangle \langle \hat a_0 \rangle \nonumber \\
& = & \frac{1}{e^{-\mu/kT} - 1} ,\label{chi2}
\end{eqnarray}
 showing the usual result that a susceptibility is always related to the fluctuations of the corresponding extensive variable. The second line in (\ref{chi2}) assumes that $\mu < 0$, that is, that the system is above BEC. This startling result indicates that the Bose susceptibility is the {\it number} of particles in the condensate, namely, $\chi = \overline n_0$. Independently of the previous comment, we emphasize that both equations, (\ref{chi1}) and (\ref{chi2}), are valid within GB treatment. But this can only be true if $\mu \to 0^-$. Thus, the Bose susceptibility can only be found, approximately, as $\chi \approx - kT/\mu$. \\

Although GB approach is of a mean-field character for $\mu \to 0$, as shown above, it is certainly a model consistent with itself, describing a phase transition with an order parameter $\Psi_0$ and its conjugate $\nu_0$. One can further follow the procedure of finding the critical exponents. It is of interest to mention that the relationship between the chemical potential and the temperature, above $T_c$, remains as given by Eq. (\ref{mu0}), namely $|\mu|^{1/2} \sim (T-T_c)$, allowing to express the behavior of the thermodynamic properties in terms of the temperature. One finds $\chi \sim (T-T_c)^{-2}$, $\nu_0 \sim \Psi_0^5$, $\Psi_0 \sim (T_c -T)^{1/2}$, and the heat capacity $C \sim |T-T_c|^{-1}$ \cite{Gunton-prl66}, which enable to identify the exponents, $\gamma_{SM} = 2$, $\delta_{SM} = 5$, $\beta_{SM}=1/2$ and $\alpha_{SM} = -1$. In a consistent fashion, the Bose susceptibility $\chi$ is related to the particle operator fluctuations, instead of the density ones, namely
\begin{eqnarray}
\chi & = & \langle \hat a_0^\dagger \hat a_0 \rangle - \langle \hat a_0^\dagger\rangle \langle \hat a_0 \rangle \nonumber \\
& = & \int G_0(\vec r) d^3 r ,\label{chi3}
\end{eqnarray}
with the particle operator correlation being,
\begin{eqnarray}
G_0(\vec r) & = &  \langle \hat \Psi^\dagger(\vec r) \hat \Psi^\dagger(0)\rangle - \langle \hat \Psi^\dagger(\vec r) \rangle \langle\hat \Psi^\dagger(0)\rangle \nonumber \\
& = & \frac{1}{h^3}  \int \frac{e^{i\vec p\cdot \vec r/\hbar} d^3p}{e^{\beta\left(\frac{p^2}{2m}-\mu\right)}-1}, \label{G0}
\end{eqnarray}
such that for $\mu \to 0^-$ these correlations turn out to be of the Orstein-Zernicke form, $G_0(r) \sim e^{-r/\xi_{SM}}/r$, yielding the exponent $\eta_{SM} = 0$, in agreement with a mean-field theory. The correlation length scales again as $\xi_{SM} \sim |T-T_c|^{-1}$ yielding $\nu_{SM} = 1$. The previous exponents are those of the Spherical Model (SM) in $d = 3$. As we summarize below, the exponents of the exact ideal BEC differ profoundly with this set.

\section{The order parameter and the exponent equalities}

In the usual BEC approach \cite{Landau1980statistical} advocated here, the condensate particle wavefunction does not appear to play  a predominant role in determining the critical properties at BEC. On the one hand, the condensate is not a superfluid, as pointed out long ago by Landau; \cite{LLII} and, on the other, from a purely thermodynamic point of view, all the statistical physics and thermodynamics of the ideal Bose gas should be entirely contained in the density matrix of the grand canonical ensemble.
From this perspective the critical properties of the BEC transition should also be related to the usual thermodynamic variables $p$, $\rho$, $S$ and $T$, whose behavior is dictated by the critical density fluctuations. Such a phase transition can be understood as one from
a normal quantum gas to a phase composed of a condensate in coexistence with a  
thermal gas at chemical potential $\mu = 0$. 
As such, as a thermodynamic consequence, it sounds reasonable that the density of particles in the condensate $\overline n_0/V$, should be identified as the order parameter of the transition, with $\overline n_0/V =0$ for $T \ge T_c$ and $\overline n_0/V \ne 0$ for  $T < T_c$.
 If one admits this identification, one can calculate the behavior of $\overline n_0/V$ in the vicinity of $T_c$, for $T < T_c$, using Eq.(\ref{condensate}), yielding,
\begin{equation}
\frac{\overline n_0}{V} \approx \frac{3}{2} \rho \left( \frac{T_c-T}{T_c} \right)
\end{equation}
which indicates that the exponent $\beta$\cite{Fisher:1967aa,ma2018modern} is given by $\beta = 1$. \\

To further justify the identification of the condensate density as the order parameter, one can use one of the most important results in the study of critical phenomena, and in physics in general, which is the fact that near a critical point, thermodynamic properties of very different physical systems behave very similarly with equal critical exponents, of which only two are independent. The others can be obtained by means of equalities among those exponents. This is the scaling hypothesis, introduced by Widom \cite{Widom:1965aa,Fisher:1967aa,ma2018modern}, and validated by the renormalization group theory developed by Wilson \cite{Wilson:1974aa,ma2018modern}. The previous theories predict the following equalities: $\alpha + 2 \beta + \gamma = 2$, $\delta = (2 - \alpha + \gamma)/2\beta$, $\gamma = (2 - \eta)\nu$ and $\alpha = 2 - d \nu$. 
The BEC exponents here presented, $\alpha = -1$, $\beta = 1$, $\gamma = 1$, $\delta =2$, $\eta = 1$ and $\nu = 1$, with $d = 3$, satisfy those relations exactly. To the best of our knowledge this is a critical universality class on its own; that is, no other known critical phenomenon shares these critical exponents. It is of pedagogical interest to realize that in $d = 3$, ideal BEC is valuable example where exact non-classical critical exponents can be calculated. In the present context is is also relevant to recall here that for the 3D XY model the critical exponents are still under debate. \cite{Campostrini:2001,Lipa:2003aa,Burovski:2006aa}.

\section{Final remarks} 

In most treatments it is always highlighted that below the transition, $T < T_c$, there appears the condensate. 
However, there is little emphasis that the thermal gas in coexistence with the condensate {\it is not} a ``normal'' gas since its chemical potential remains zero throughout, until Absolute Zero is attained. It is more than of historical interest to point out that Einstein paid more attention to the  thermal gas than to the condensate itself.\cite{Einstein1925} The thermal gas at $T < T_c$, having $\mu =0$, has the peculiarity that its number of particles $N$ is no longer an independent thermodynamic variable, but it is rather determined by the thermodynamic state itself. That is, the only thermodynamic independent variables are the temperature $T$ (or entropy $S$) and the volume $V$ (or the pressure $p$).  Hence, at the same time that the pressure is only a function of temperature, $p = p(T)$, there is only a single stability condition established by the requirement that the heat capacity must be positive $C_V > 0$. Therefore, it is meaningless to calculate the isothermal compressibility $\kappa_T$: it is not only not defined but also it does not indicate anything regarding the stability of the gas. The stability of the thermal gas in coexistence with the condensate is analogous to the stability of a gas of photons: both have $\mu = 0$ and, therefore, the entropy $S$ depends on the internal energy $E$ and on the volume $V$, but not on the number of particles $N$; the unique stability condition is then $(\partial^2 S/\partial E^2)_V < 0$. As far as the density correlations are concerned, it is certainly interesting to realize that the correlation function has no characteristic length, since $\xi$ is infinite for $\mu = 0$,  and so, the thermal gas in coexistence with the condensate is always in a ``critical" state in the sense that its correlations are long range decaying algebraically $\sim r^{-2}$ for all temperatures $T \le T_c$. The thermal part of the gas still shows critical exponents $\alpha = -1$ and $\eta = 1$, but the other exponents, $\gamma$, $\nu$ and $\delta$ apparently cannot be defined. In this regard, one could still argue that the condensate density $\overline n_0/V$ is not necessarily the order parameter and, thus, that the exponent $\beta$ may not be equal to one. 
%
%
%
We believe it would be of interest to perform a renormalization group approach in order to elucidate the general characteristics of this ``new'' universality class.\\

We acknowledge support from grants CONACYT (Mexico) 255573 and PAPIIT-IN105217 (UNAM).
 I.R.A. thanks CONACYT for a graduate studies scholarship, and F.J.P.C. thanks the C\'atedras CONACYT program.

\bibliographystyle{apsrev4-1}
\bibliography{References}

\end{document}